\pdfoutput=1
\documentclass[12pt,a4paper]{article}

\usepackage{ifthen} 
\newboolean{pdflatex}
\setboolean{pdflatex}{true} 

\newboolean{articletitles}
\setboolean{articletitles}{true} 

\newboolean{uprightparticles}
\setboolean{uprightparticles}{false} 

\usepackage{booktabs}

\def\paperauthors{LHCb collaboration} 

\def\paperasciititle{Measurement of the Omegac0 and Xic0 lifetimes using hadronic b-baryon decays} 
\def\papertitle{Measurement of the \Omegac and \Xicz baryon lifetimes using hadronic \bquark-baryon decays} 
\def\paperkeywords{{High Energy Physics}, {LHCb}, {Hadron spectroscopy}, {Lifetime measurements}} 
\def\papercopyright{\the\year\ CERN for the benefit of the LHCb collaboration} 
\def\paperlicence{CC BY 4.0 licence}
\def\paperlicenceurl{https://creativecommons.org/licenses/by/4.0/}

\newif\ifEnableSectionTOCLinks
\EnableSectionTOCLinksfalse 

\usepackage[top=1in, bottom=1.25in, left=1in, right=1in]{geometry}

\usepackage{microtype}
\usepackage{lineno}  
\usepackage{xspace} 
\usepackage{caption} 

\usepackage{graphicx}  
\usepackage{color}
\usepackage{colortbl}
\graphicspath{{./figs/}} 

\usepackage{amsmath} 
\usepackage{amssymb}
\usepackage{amsfonts}
\usepackage{upgreek} 

\newcommand*\patchAmsMathEnvironmentForLineno[1]{%
\expandafter\let\csname old#1\expandafter\endcsname\csname #1\endcsname
\expandafter\let\csname oldend#1\expandafter\endcsname\csname
end#1\endcsname
 \renewenvironment{#1}%
   {\linenomath\csname old#1\endcsname}%
   {\csname oldend#1\endcsname\endlinenomath}%
}
\newcommand*\patchBothAmsMathEnvironmentsForLineno[1]{%
  \patchAmsMathEnvironmentForLineno{#1}%
  \patchAmsMathEnvironmentForLineno{#1*}%
}
\AtBeginDocument{%
\patchBothAmsMathEnvironmentsForLineno{equation}%
\patchBothAmsMathEnvironmentsForLineno{align}%
\patchBothAmsMathEnvironmentsForLineno{flalign}%
\patchBothAmsMathEnvironmentsForLineno{alignat}%
\patchBothAmsMathEnvironmentsForLineno{gather}%
\patchBothAmsMathEnvironmentsForLineno{multline}%
\patchBothAmsMathEnvironmentsForLineno{eqnarray}%
}

\usepackage[pdftex,
            pdfauthor={\paperauthors},
            pdftitle={\paperasciititle},
            pdfkeywords={\paperkeywords}]{hyperref}
\usepackage{hyperxmp}
\hypersetup{
    pdfcopyright={Copyright (C) \papercopyright},
    pdflicenseurl={\paperlicenceurl}
}

\usepackage[colorinlistoftodos,textsize=scriptsize]{todonotes}

\usepackage[bottom,flushmargin,hang,multiple]{footmisc}

\usepackage[all]{hypcap} 

\usepackage{xspace} 
\usepackage{upgreek}

\def\lhcb   {\mbox{LHCb}\xspace}

\def\belletwo {\mbox{Belle~II}\xspace}

\def\MagUp {\mbox{\em Mag\kern -0.05em Up}\xspace}

\ifthenelse{\boolean{uprightparticles}}%
{
 
 \def\Pgamma      {\ensuremath{\upgamma}\xspace}

 \def\Ppi         {\ensuremath{\uppi}\xspace}

 \def\PDelta      {\ensuremath{\Delta}\xspace}                 
 \def\PXi         {\ensuremath{\Xi}\xspace}                 
 \def\PLambda     {\ensuremath{\Lambda}\xspace}                 
 \def\PSigma      {\ensuremath{\Sigma}\xspace}                 
 \def\POmega      {\ensuremath{\Omega}\xspace}                 
 \def\PUpsilon    {\ensuremath{\Upsilon}\xspace}
 \let\oldPi\Pi
 \def\PPi         {\ensuremath{\oldPi}\xspace}

 \def\PB      {\ensuremath{\mathrm{B}}\xspace}                 
 \def\PD      {\ensuremath{\mathrm{D}}\xspace}                 

 \def\PK      {\ensuremath{\mathrm{K}}\xspace}                 
 \def\PW      {\ensuremath{\mathrm{W}}\xspace}                 
 \def\Pb      {\ensuremath{\mathrm{b}}\xspace}                 
 \def\Pc      {\ensuremath{\mathrm{c}}\xspace}                 
 \def\Pd      {\ensuremath{\mathrm{d}}\xspace}                 

 \def\Pp      {\ensuremath{\mathrm{p}}\xspace}                 

 \def\Ps      {\ensuremath{\mathrm{s}}\xspace}

 \def\thebaroffset{0.0em}
}
{
 
 \def\Pgamma      {\ensuremath{\gamma}\xspace}

 \def\Ppi         {\ensuremath{\pi}\xspace}

 \mathchardef\PDelta="7101
 \mathchardef\PXi="7104
 \mathchardef\PLambda="7103
 \mathchardef\PSigma="7106
 \mathchardef\POmega="710A
 \mathchardef\PUpsilon="7107
 \mathchardef\PPi="7105
 \def\PB      {\ensuremath{B}\xspace}                 
 \def\PD      {\ensuremath{D}\xspace}                 

 \def\PK      {\ensuremath{K}\xspace}                 
 \def\PW      {\ensuremath{W}\xspace}                 
 \def\Pb      {\ensuremath{b}\xspace}                 
 \def\Pc      {\ensuremath{c}\xspace}                 
 \def\Pd      {\ensuremath{d}\xspace}                 

 \def\Pp      {\ensuremath{p}\xspace}                 

 \def\Ps      {\ensuremath{s}\xspace}

 \def\thebaroffset{0.18em}
}
\newcommand{\offsetoverline}[2][\thebaroffset]{\kern #1\overline{\kern -#1 #2}}%

\makeatletter
\ifcase \@ptsize \relax
  \newcommand{\miniscule}{\@setfontsize\miniscule{4}{5}}
\or
  \newcommand{\miniscule}{\@setfontsize\miniscule{5}{6}}
\or
  \newcommand{\miniscule}{\@setfontsize\miniscule{5}{6}}
\fi
\makeatother

\DeclareRobustCommand{\optbar}[1]{\shortstack{{\miniscule (\rule[.5ex]{1.25em}{.18mm})}
  \\ [-.7ex] $#1$}}

\def\g      {{\ensuremath{\Pgamma}}\xspace}

\def\dquark    {{\ensuremath{\Pd}}\xspace}

\def\squark    {{\ensuremath{\Ps}}\xspace}

\def\cquark    {{\ensuremath{\Pc}}\xspace}

\def\bquark    {{\ensuremath{\Pb}}\xspace}

\def\pion   {{\ensuremath{\Ppi}}\xspace}
\def\piz    {{\ensuremath{\pion^0}}\xspace}
\def\pip    {{\ensuremath{\pion^+}}\xspace}
\def\pim    {{\ensuremath{\pion^-}}\xspace}

\def\kaon    {{\ensuremath{\PK}}\xspace}

\def\KorKbar {\kern \thebaroffset\optbar{\kern -\thebaroffset \PK}{}\xspace}

\def\Kp      {{\ensuremath{\kaon^+}}\xspace}
\def\Km      {{\ensuremath{\kaon^-}}\xspace}

\def\D       {{\ensuremath{\PD}}\xspace}

\def\DorDbar {\kern \thebaroffset\optbar{\kern -\thebaroffset \PD}\xspace}
\def\Dz      {{\ensuremath{\D^0}}\xspace}

\def\Dp      {{\ensuremath{\D^+}}\xspace}
\def\Dm      {{\ensuremath{\D^-}}\xspace}

\def\DpDm    {\ensuremath{\Dp {\kern -0.16em \Dm}}\xspace}

\def\Dstarz  {{\ensuremath{\D^{*0}}}\xspace}

\def\B       {{\ensuremath{\PB}}\xspace}

\def\BorBbar {\kern \thebaroffset\optbar{\kern -\thebaroffset \PB}\xspace}

\def\Bd      {{\ensuremath{\B^0}}\xspace}

\def\BdorBdbar {\kern \thebaroffset\optbar{\kern -\thebaroffset \Bd}\xspace}

\def\Bub     {{\ensuremath{\B^-}}\xspace}

\def\Bs      {{\ensuremath{\B^0_\squark}}\xspace}

\def\BsorBsbar {\kern \thebaroffset\optbar{\kern -\thebaroffset \Bs}\xspace}

\def\Y#1S{\ensuremath{\PUpsilon{(#1S)}}\xspace}

\def\proton      {{\ensuremath{\Pp}}\xspace}

\def\Lz          {{\ensuremath{\PLambda}}\xspace}

\def\LorLbar     {\kern \thebaroffset\optbar{\kern -\thebaroffset \PLambda}\xspace}

\def\Xires       {{\ensuremath{\PXi}}\xspace}

\def\Omegares    {{\ensuremath{\POmega}}\xspace}

\def\Lc          {{\ensuremath{\Lz^+_\cquark}}\xspace}

\def\Xic         {{\ensuremath{\Xires_\cquark}}\xspace}
\def\Xicz        {{\ensuremath{\Xires^0_\cquark}}\xspace}
\def\Xicp        {{\ensuremath{\Xires^+_\cquark}}\xspace}

\def\Omegac      {{\ensuremath{\Omegares^0_\cquark}}\xspace}

\def\Xibm         {{\ensuremath{\Xires^-_\bquark}}\xspace}

\def\Omegab       {{\ensuremath{\Omegares^-_\bquark}}\xspace}

\newcommand{\decay}[2]{\mbox{\ensuremath{#1\!\to #2}}\xspace} 

\def\to                 {\ensuremath{\rightarrow}\xspace}

\newcommand{\tauDz}{{\ensuremath{\tau_{\Dz}}}\xspace}

\def\grpsuthree {{\ensuremath{\mathrm{SU}(3)}}\xspace}

\def\AT#1     {\ensuremath{A_{\mathrm{T}}^{#1}}\xspace}           

\def\C#1      {\ensuremath{\mathcal{C}_{#1}}\xspace}                       
\def\Cp#1     {\ensuremath{\mathcal{C}_{#1}^{'}}\xspace}                    
\def\Ceff#1   {\ensuremath{\mathcal{C}_{#1}^{\mathrm{(eff)}}}\xspace}        
\def\Cpeff#1  {\ensuremath{\mathcal{C}_{#1}^{'\mathrm{(eff)}}}\xspace}       
\def\Ope#1    {\ensuremath{\mathcal{O}_{#1}}\xspace}                       
\def\Opep#1   {\ensuremath{\mathcal{O}_{#1}^{'}}\xspace}                    

\newcommand{\nospaceunit}[1]{\ensuremath{\text{#1}}}       
\newcommand{\aunit}[1]{\ensuremath{\text{\,#1}}}       

\newcommand{\tev}{\aunit{Te\kern -0.1em V}\xspace}
\newcommand{\gev}{\aunit{Ge\kern -0.1em V}\xspace}
\newcommand{\mev}{\aunit{Me\kern -0.1em V}\xspace}
\newcommand{\kev}{\aunit{ke\kern -0.1em V}\xspace}
\newcommand{\ev}{\aunit{e\kern -0.1em V}\xspace}
 
\newcommand{\mevc}{\ensuremath{\aunit{Me\kern -0.1em V\!/}c}\xspace}
\newcommand{\gevc}{\ensuremath{\aunit{Ge\kern -0.1em V\!/}c}\xspace}
\newcommand{\mevcc}{\ensuremath{\aunit{Me\kern -0.1em V\!/}c^2}\xspace}
\newcommand{\gevcc}{\ensuremath{\aunit{Ge\kern -0.1em V\!/}c^2}\xspace}

\def\mum  {\ensuremath{\,\upmu\nospaceunit{m}}\xspace}

\def\fb   {\ensuremath{\aunit{fb}}\xspace}
\def\invfb   {\ensuremath{\fb^{-1}}\xspace}

\def\ps   {\ensuremath{\aunit{ps}}\xspace}
\def\fs   {\aunit{fs}}

\newcommand{\stat}{\aunit{(stat)}\xspace}
\newcommand{\syst}{\aunit{(syst)}\xspace}

\newcommand{\chisq}{\ensuremath{\chi^2}\xspace}
\newcommand{\chisqndf}{\ensuremath{\chi^2/\mathrm{ndf}}\xspace}
\newcommand{\chisqip}{\ensuremath{\chi^2_{\text{IP}}}\xspace}

\def\deriv {\ensuremath{\mathrm{d}}}

\def\gsim{{~\raise.15em\hbox{$>$}\kern-.85em
          \lower.35em\hbox{$\sim$}~}\xspace}
\def\lsim{{~\raise.15em\hbox{$<$}\kern-.85em
          \lower.35em\hbox{$\sim$}~}\xspace}

\def\sPlot{\mbox{\em sPlot}\xspace}

\def\sqs   {\ensuremath{\protect\sqrt{s}}\xspace}

\def\pt         {\ensuremath{p_{\mathrm{T}}}\xspace}

\def\ptot       {\ensuremath{p}\xspace}

\def\evtgen     {\mbox{\textsc{EvtGen}}\xspace}

\def\geant      {\mbox{\textsc{Geant4}}\xspace}

\def\photos     {\mbox{\textsc{Photos}}\xspace}

\def\pythia     {\mbox{\textsc{Pythia}}\xspace}

\def\tell1  {TELL1\xspace}
\def\ukl1   {UKL1\xspace}

\newcommand{\lhcborcid}[1]{\href{https://orcid.org/#1}{\hspace*{0.1em}\raisebox{-0.45ex}{\includegraphics[width=1em]{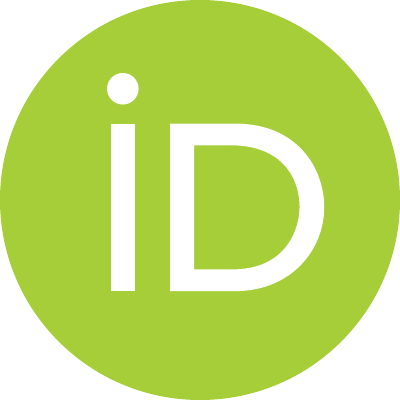}}}}

\newcommand{\tauOmc}{{\ensuremath{\tau_{\Omegac}}}\xspace}
\newcommand{\tauXicz}{{\ensuremath{\tau_{\Xicz}}}\xspace}
\newcommand{\tauLc}{{\ensuremath{\tau_{\Lc}}}\xspace}
\newcommand{\tauXicp}{{\ensuremath{\tau_{\Xicp}}}\xspace}

\def\OmToh        {\decay{\Omegab}{\Omegac(\decay{}{\proton\Km\Km\pip})\,\pim}}
\def\XiToh        {\decay{\Xibm}{\Xicz(\decay{}{\proton\Km\Km\pip})\,\pim}}
\def\bToDh        {\decay{\Bub}{\Dz(\decay{}{\Km\Kp\pim\pip})\,\pim}}

\def\OmcTohhhh    {\decay{\Omegac}{\proton\Km\Km\pip}}
\def\XicTohhhh    {\decay{\Xicz}{\proton\Km\Km\pip}}
\def\DzTohhhh     {\decay{\Dz}{\Km\Kp\pim\pip}}

\def\Oms          {\decay{\Omegab}{\Omegac\pim}}
\def\Xis          {\decay{\Xibm}{\Xicz\pim}}
\def\bees         {\decay{\Bub}{\Dz\pim}}
\def\OmsCS        {\decay{\Omegab}{\Omegac\Km}}
\def\XisCS        {\decay{\Xibm}{\Xicz\Km}}
\def\beesCS       {\decay{\Bub}{\Dz\Km}}

\def\pKKpi           {{\ensuremath{\proton\Km\Km\pip}}\xspace}
\def\KKpipi          {{\ensuremath{\Kp\Km\pip\pim}}\xspace}
\def\Xicpi           {{\ensuremath{\Xicz\pim}}\xspace}
\def\Omcpi           {{\ensuremath{\Omegac\pim}}\xspace}
\def\Dzpi            {{\ensuremath{\Dz\pim}}\xspace}
\def\Omegacstar      {{\ensuremath{\Omegares_{\cquark}(2770)^{0}}}\xspace}
\def\Xiczprime       {{\ensuremath{\Xires^{\prime 0}_{\cquark}}}\xspace}

\def\OmTostarfull    {\decay{\Omegab}{\Omegacstar(\decay{}{\Omegac\g})\,\pim}}

\def\OmbToOmcpipizShort    {\decay{\Omegab}{\Omegac\pim\piz}}

\def\XiToOmczPim     {\decay{\Xibm}{\Omegac\pim}}

\def\XiToXistar      {\decay{\Xibm}{\Xic(2645)^{0}(\decay{}{\Xicz \piz})\,\pim}}

\def\XibToXicpipizShort    {\decay{\Xibm}{\Xicz\pim\piz}}

\def\XiToprime        {\decay{\Xibm}{\Xiczprime(\decay{}{\Xicz\g})\,\pim}}

\def\bTogamma        {\decay{\Bub}{\Dstarz(\decay{}{\Dz\gamma})\pim}}

\def\sPlot{\mbox{\em sPlot}\xspace}

\newcommand{\chisqdtf}{\ensuremath{\chi^2_{\text{DTF}}/\mathrm{ndf}}\xspace}

\def\yieldOmc {355 \pm 26}

\def\yieldXicRound   {8\,260 \pm 100}
\def\yieldDzOmRound  {53\,420 \pm 260}
\def\yieldDzXiRound  {48\,720 \pm 250}

\newcommand{\dzUncert}{\aunit{(\tauDz)}\xspace}

\def\systvalueOmStatSWeights {265.5 \pm 19.8 \stat}

\def\systvalueOmStatRatioofYields {276.3 \pm 19.4 \stat}
\def\systvalueOmStatSystRatioofYields {\systvalueOmStatRatioofYields \pm 1.8 \syst}
\def\systvalueOmroy {\systvalueOmStatSystRatioofYields \pm 0.7\dzUncert}

\def\systvalueXiStatSWeights {148.2 \pm 2.6\stat}

\def\systvalueXiStatRatioofYields {149.2 \pm~\,2.5\stat}
\def\systvalueXiStatSystRatioofYields {\systvalueXiStatRatioofYields \pm 0.9\syst}
\def\systvalueXiroy {\systvalueXiStatSystRatioofYields \pm 0.4\dzUncert}

\def\averageOmc{274.8 \pm 10.5}
\def\averageXic{150.7 \pm~\,1.6}

\hypersetup{
  colorlinks   = true, 
  urlcolor     = blue, 
  linkcolor    = blue, 
  citecolor    = red   
}

\ifEnableSectionTOCLinks
    \usepackage[explicit]{titlesec} 
    
    \let\oldcontentsline\contentsline
    \renewcommand

    \titleformat{\section}{\normalfont\Large\bf}{\hyperlink{tocsection.\thesection}{{\thesection} \parbox[t]{\dimexpr\textwidth-1pc}{#1}}}{1pc}{}

    \titleformat{\subsection}{\normalfont\bf}{\hyperlink{tocsubsection.\thesubsection}{{\thesubsection} \parbox[t]{\dimexpr\textwidth-1pc}{#1}}}{1pc}{}

    \titleformat{name=\section,numberless}[display]{}{}{0pt}{\normalfont\Huge\bfseries #1}
\fi

\usepackage{cite} 
\usepackage{mciteplus}

\begin{document}

\renewcommand{\thefootnote}{\fnsymbol{footnote}}
\setcounter{footnote}{1}


\begin{titlepage}
\pagenumbering{roman}

\vspace*{-1.5cm}
\centerline{\large EUROPEAN ORGANIZATION FOR NUCLEAR RESEARCH (CERN)}
\vspace*{1.5cm}
\noindent
\begin{tabular*}{\linewidth}{lc@{\extracolsep{\fill}}r@{\extracolsep{0pt}}}
\vspace*{-1.5cm}\mbox{\!\!\!\includegraphics[width=.14\textwidth]{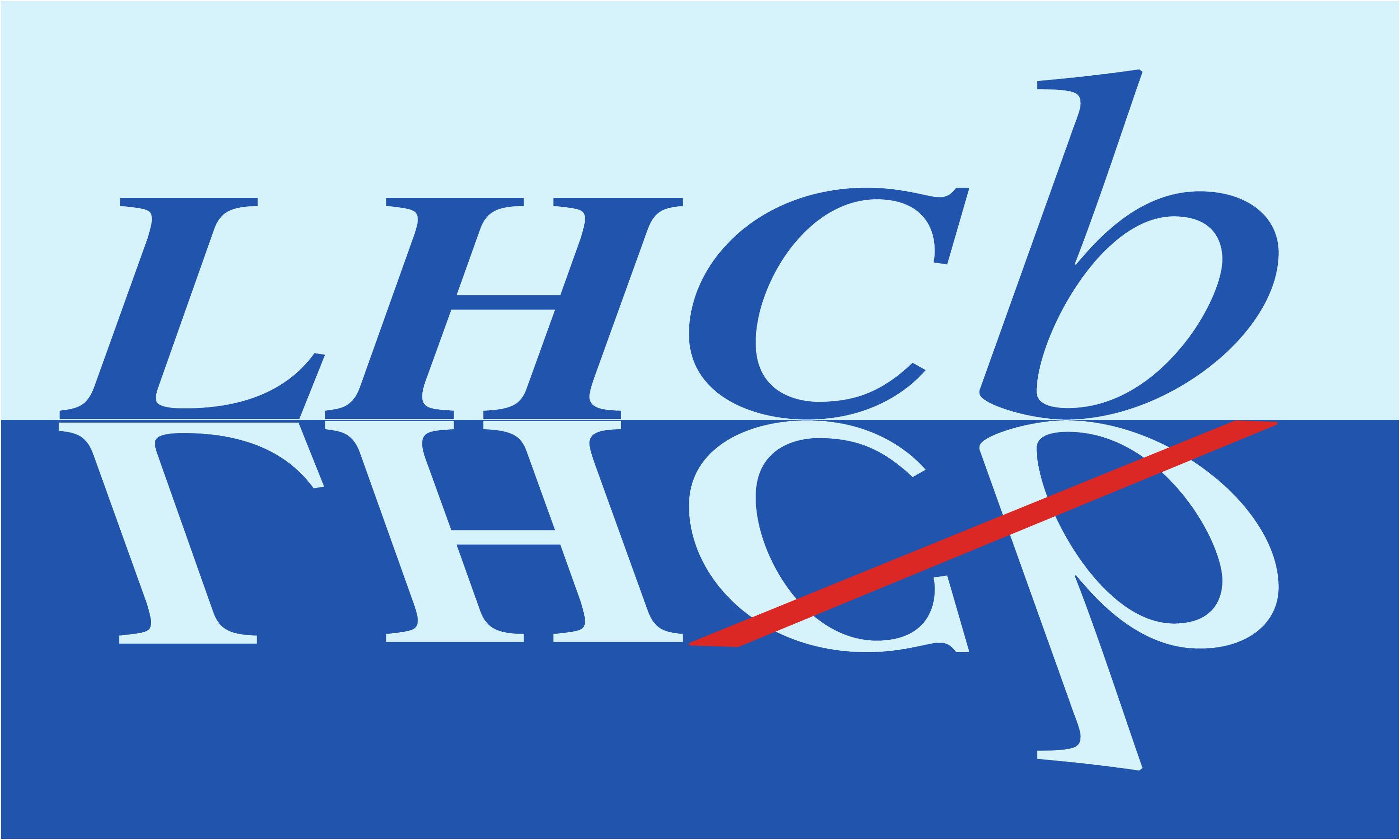}} & &
\\
 & & CERN-EP-2025-117 \\  
 & & LHCb-PAPER-2025-013 \\  
 & & 1 October, 2025 \\ 
 & & \\
\end{tabular*}

\vspace*{2.0cm}

{\normalfont\bfseries\boldmath\huge
\begin{center}
  \papertitle 
\end{center}
}

\vspace*{1.0cm}

\begin{center}
\paperauthors\footnote{Authors are listed at the end of this paper.}
\end{center}

\vspace{\fill}

\begin{abstract}
  \noindent
  The lifetimes of the $\Omegac$ and $\Xicz$ baryons are measured using a $pp$ collision dataset collected by the $\lhcb$ experiment, corresponding to an integrated luminosity of $9\invfb$. The charm baryons are produced in the fully reconstructed decay chains $\OmToh$ and $\XiToh$. The measurement uses topologically and kinematically similar $\bToDh$ decays for normalisation. The measured lifetimes are 
  \begin{eqnarray*}
    \tauOmc \!\! &=& \!\!\systvalueOmroy\fs,\\ 
    \tauXicz \!\! &=&\!\! \systvalueXiroy\fs,  
  \end{eqnarray*}
  where the first uncertainty is statistical, the second systematic and the third due to the uncertainty of the $\Dz$ lifetime. These results are consistent with previous measurements performed by the $\lhcb$ experiment.
\end{abstract}

\vspace*{2.0cm}

\begin{center}
  Published in JHEP 09 (2025) 157
\end{center}

\vspace{\fill}

{\footnotesize 
\centerline{\copyright~\papercopyright. \href{\paperlicenceurl}{\paperlicence}.}}
\vspace*{2mm}

\end{titlepage}


\newpage
\setcounter{page}{2}
\mbox{~}


\renewcommand{\thefootnote}{\arabic{footnote}}
\setcounter{footnote}{0}


\cleardoublepage


\pagestyle{plain} 
\setcounter{page}{1}
\pagenumbering{arabic}



\section{Introduction}
\label{sec:introduction}


Charm and beauty hadron lifetime measurements are important means to explore and document the properties of hadrons. Experimentally, accurate simulation of particle decays requires good knowledge of the lifetimes. From a theoretical viewpoint, charm hadron lifetimes are an excellent testing ground to evaluate different techniques for calculating hadron properties. Recent measurements of the \Omegac and \Xicz lifetimes are in tension with earlier results, and this paper presents new results that confirm the revised lifetimes.


The lifetime of the \Omegac baryon, with a valence quark content of \cquark\squark\squark, was first measured by the E687 and WA89 collaborations~\cite{E687:1995cvt,WA89:1995lbz} and the FOCUS experiment~\cite{FOCUS:2003ylx}, resulting in an average of $69\pm12\fs$~\cite{PDG2018}. In a more recent measurement, the \lhcb experiment measured the \Omegac lifetime to be almost four times larger than, and inconsistent with, the previous average. This measurement was based on a significantly larger sample of \Omegac baryons originating from semileptonic \bquark-baryon decays~\cite{LHCb-PAPER-2018-028}. This result was subsequently confirmed by a measurement from the \lhcb experiment using an independent sample of \Omegac baryons produced promptly in $pp$ collisions~\cite{LHCB-PAPER-2021-021}, yielding an average for the \lhcb measurements of $275\pm12\fs$. The \belletwo experiment recently measured the \Omegac lifetime to be $243\pm49$\fs~\cite{PhysRevD.107.L031103}, consistent with the longer lifetime reported by \lhcb.

Similarly, the lifetime of the \Xicz baryon, with a valence quark content of \cquark\squark\dquark, was first measured by the ACCMOR, WA687 and FOCUS collaborations~\cite{ACCMOR:1990qio,E687:1992jlh,Focus:2002vun}, resulting in an average of $112^{+13}_{-10}\fs$\cite{PDG2018}. The more recent measurements by the \lhcb experiment, using \Xicz baryons produced in semileptonic decays~\cite{LHCb-PAPER-2019-008} and promptly in $pp$ collisions~\cite{LHCB-PAPER-2021-021}, average to a lifetime of $152.0\pm2.0\fs$, in tension with the previous average. These discrepancies in measured lifetimes for the \Omegac and \Xicz baryons highlight the need for additional measurements.


The lifetimes of charm baryons can be calculated using heavy-quark-expansion theory~\cite{Khoze:1983yp,Bigi:1991ir,Bigi:1992su,Blok:1992hw,*Blok:1992he,
Neubert:1997gu,Uraltsev:1998bk,Bigi:1995jr}, which is an effective theory based on an expansion in inverse powers of the mass of the heaviest quark. At leading order, the lifetime only depends on the mass of the charm quark, resulting in the same lifetime for all four weakly decaying single-charm baryons. Differences in the lifetimes of the single-charm baryons are expected to arise from higher-order effects, such as weak $\PW$-exchange and Pauli interference, caused by the spectator quarks. Charm hadron lifetimes are more sensitive to these contributions than beauty hadrons since the size of the contributions increases as the mass of the heaviest quark decreases~\cite{Kirk:2017juj, Lenz:2013aua,cheng2021charmed,Bianco:2003vb,Bellini:1996ra,Blok:1991st}.

The higher-order effects on the lifetimes of single-charm baryons have been evaluated to give a hierarchy of \mbox{$\tauXicp>\tauLc>\tauXicz>\tauOmc$}~\cite{Blok:1991st,Bellini:1996ra,Cheng:1997xba,Bianco:2003vb,cheng2021charmed}, consistent with the earlier measurements of the lifetimes. Different treatments of the higher-order effects predict the hierarchy \mbox{$\tauXicp>\tauOmc>\tauLc>\tauXicz$}~\cite{Blok:1991st,Cheng:2018rkz,cheng2021charmed}, which is consistent with the more recent measurements. Hence, experimental input is crucial to arbitrate between these different calculations. 


The measurements presented in this paper exploit charm baryons produced in hadronic \mbox{\bquark-baryon} decays recorded by the \lhcb experiment, specifically \OmToh and \XiToh decays.\footnote{The inclusion of charge-conjugate processes is implied throughout.} These fully reconstructed decays have well defined production and decay vertices and are statistically independent from previous measurements by the \lhcb experiment. The lifetimes are measured using the topologically and kinematically similar \bToDh decay for normalisation to minimise the uncertainties associated with the decay time acceptance. Two different methods for the background subtraction are used to cross-check the results, giving consistent values. In order to avoid experimenter's bias, the results of the analysis were not examined until the full procedure had been finalised.

The data sample was collected by the \lhcb experiment in the years 2011, 2012 and $2015-2018$ at centre-of-mass energies of \sqs = 7, 8 and 13\tev, respectively, corresponding to an integrated luminosity of 9\invfb.


\section{Detector and simulation}
\label{sec:detector}


The \lhcb detector~\cite{LHCb-DP-2008-001,LHCb-DP-2014-002} is a single-arm forward spectrometer covering the \mbox{pseudorapidity} range $2<\eta <5$, designed for the study of particles containing \bquark or \cquark quarks. The detector includes a high-precision tracking system consisting of a silicon-strip vertex detector surrounding the $pp$ interaction region~\cite{LHCb-DP-2014-001}, a large-area silicon-strip detector located upstream of a dipole magnet with a bending power of about $4{\mathrm{\,T\,m}}$, and three stations of silicon-strip detectors and straw
drift tubes~\cite{LHCb-DP-2013-003,LHCb-DP-2017-001} placed downstream of the magnet. The tracking system provides a measurement of the momentum, \ptot, of charged particles with a relative uncertainty that varies from 0.5\% at low momentum to 1.0\% at 200\gevc. The minimum distance of a track to a primary $pp$ collision vertex (PV), the impact parameter (IP), is measured with a resolution of $(15+29/\pt)\mum$, where \pt is the component of the momentum transverse to the beam, in\,\gevc. Different types of charged hadrons are distinguished using information from two ring-imaging Cherenkov detectors~\cite{LHCb-DP-2012-003}.  Photons, electrons and hadrons are identified by a calorimeter system consisting of
scintillating-pad and preshower detectors, an electromagnetic and a hadronic calorimeter. Muons are identified by a system composed of alternating layers of iron and multiwire proportional chambers~\cite{LHCb-DP-2012-002}.


The online event selection is performed by a trigger~\cite{LHCb-DP-2012-004},  which consists of a hardware stage, based on information from the calorimeter and muon systems, followed by a software stage, which applies a full event reconstruction. 
At the hardware trigger stage, events are required to have a muon with high \pt or a hadron, photon or electron with high transverse energy in the calorimeters. The signal candidates selected for the measurement are not required to have passed a specific set of criteria at the hardware stage; all candidates accepted by the trigger are included. 
The software trigger requires a two-, three- or four-track secondary vertex with a significant displacement from any PV. At least one charged particle must have a transverse momentum $\pt > 1.6\gevc$ and be inconsistent with originating from a PV. A multivariate algorithm~\cite{BBDT,LHCb-PROC-2015-018} is used for the identification of secondary vertices consistent with the decay of a beauty hadron.


Simulation is required to determine how the detector acceptance and selection criteria affect the decay time distributions, and to determine the decay time resolution. In addition, simulation is used to develop the candidate selections. In the simulation, $pp$ collisions are generated using~\pythia~\cite{Sjostrand:2007gs,Sjostrand:2006za}  with a specific \lhcb configuration~\cite{LHCb-PROC-2010-056}. Decays of unstable particles are described by \evtgen~\cite{Lange:2001uf}, in which final-state radiation is generated using \photos~\cite{davidson2015photos}. The interaction of the generated particles with the detector, and its response, are implemented using the \geant toolkit~\cite{Allison:2006ve, *Agostinelli:2002hh} as described in Ref.~\cite{LHCb-PROC-2011-006}. To make more efficient use of computing resources, the underlying $pp$ interaction is reused multiple times and combined with independently generated signal decays each time~\cite{LHCb-DP-2018-004}. The simulated data are produced with charm and beauty hadron lifetimes consistent with the known values~\cite{PDG2024}, or weighted to match those values if different.


The simulation provides a good description of the data, with the agreement further improved by data-driven corrections. The \bquark-baryon \pt spectrum, track multiplicity and hardware-trigger efficiency are corrected with weights determined from the observed differences between data and simulation. The tracking efficiency corrections are applied using tables binned in pseudorapidity and \pt, derived from calibration samples. The simulation is corrected so that the fractions of candidates triggered by the signal or by other particles in the event match those in data. The particle identification performance in simulation is corrected using information from calibration data samples~\cite{LHCb-PUB-2016-021,LHCb-DP-2018-001}.


The lifetime is measured from the distribution of decay times $t$ for the signal candidates, evaluated in the particles' rest frames. It is determined from the reconstructed decay length $l$, invariant mass $m$ and momentum $p$ of the signal candidate, where $t=\frac{m\cdot l}{p}$~\cite{LHCb-DP-2014-001}. This relation is evaluated in a fit where the candidate is constrained to originate from its associated PV. The associated PV is that to which the candidates has the smallest \chisqip, defined as the difference in the vertex-fit \chisq of a given PV reconstructed with and without the track under consideration.


\section{Selection of signal candidates}
\label{sec:selection}

The selection of signal candidates exploits the long lifetime of the \bquark-hadrons, avoiding requirements on quantities that directly affect the decay time distribution of the charm hadrons. Hence, the selection efficiencies only depend weakly on the charm hadron decay time, which simplifies the corrections for the decay time acceptance. The selection of the normalisation mode closely matches that of the signal modes, to ensure good cancellation of the acceptance functions in the ratio of the signal and normalisation modes. Differences are limited to particle identification requirements and mass-range selections.


Four charged tracks, not originating from a PV and with requirements on their \ptot and \pt, are selected to form a good-quality displaced vertex. The tracks must be consistent with the pion, kaon or proton hypotheses, as required to form the \OmcTohhhh, \XicTohhhh and \DzTohhhh candidate decays.

Each charm hadron candidate is combined with a charged track that does not originate from a PV, has a high probability of being a pion and satisfies requirements on its \ptot and \pt, to form a beauty-hadron candidate. The charm hadron and the additional track are required to originate from a common vertex, which must be displaced from all PVs and be of good quality. In addition, the reconstructed momentum vector of the \bquark-hadron candidate must point back to the associated PV. 
A kinematic fit to the full decay tree~\cite{Hulsbergen:2005pu} is performed on these \OmToh, \XiToh and \bToDh candidates, in which they are constrained to originate from their associated PV. A loose requirement on the quality of that fit is applied, estimated by its \chisq per degree of freedom (\chisqdtf). 


A boosted decision tree (BDT) classifier~\cite{Breiman,AdaBoost}, implemented in the TMVA toolkit~\cite{Hocker:2007ht}, is employed to further reject background. The classifier is trained using proxy datasets to represent signal and background decays. The signal is represented by simulated signal decays. The background is represented by candidates from data with \pKKpi (\KKpipi) invariant masses outside the charm baryon (\Dz meson) signal regions, or with \Omegac\pim, \Xicz\pim or \Dz\pim invariant masses significantly displaced above the \bquark-hadron signal peaks. These training datasets are subject to the selection criteria described above.

Fifteen variables are used in the BDT to differentiate between signal and background. Among these, the variables associated with the pions from the \bquark-hadron decays are their \pt, their \chisqip and their probability of being a pion. Variables associated with the charm hadrons are their total momentum, the sum of \pt of their decay products and the probability of each decay product being assigned the correct particle identity. Variables associated with the \bquark-hadron candidates are their \pt, \chisqip and \chisqdtf, the \chisqndf of their decay vertex fits, the angle between their reconstructed flight path and momentum vector, and the flight distance significance, defined as the distance divided by its uncertainty.

The requirements on the BDT output variable for accepting charm baryon candidates are optimised by maximising the signal significance, which is defined as $S/\sqrt{S+B}= S_0 \epsilon_s/\sqrt{S_0\epsilon_s+B}$. The background yield in the signal region, $B$, is determined from a fit to data. The signal yield $S_0$ is determined from a fit to data when applying a loose BDT requirement, which is then scaled with the change in selection efficiency ($\epsilon_s$) determined from simulation, as the BDT requirement is tightened. This ensures that the measurement is not biased by the optimisation procedure. For the normalisation channel, a requirement that is tighter than the optimum given by the described procedure is applied. The number of selected candidates in the normalisation mode is significantly higher than for the signal modes; hence, the tighter selection does not affect the statistical precision, and the very clean selected sample simplifies the modelling of the invariant mass spectrum. 

The training of the BDT classifier and the optimisation of the output variable requirement are performed separately for the data-taking periods 2011\,--\,2012 and 2015\,--\,2018, to account for the differences in running conditions and centre-of-mass energies.


Candidates are required to have \pKKpi (\KKpipi) invariant masses within three times the mass resolution around the known charm baryon (\Dz) masses~\cite{PDG2024}, and \Omcpi, \Xicpi and \Dzpi invariant masses in ranges from 300, 200 and 150\mevcc below to 550, 550 and 450\mevcc above the known \Omegab, \Xibm and \Bub masses, respectively.

The selected decay time ranges for the two charm baryons are chosen to extend up to six times their lifetimes, which gives ranges of $-0.15$ to $1.50\ps$ for the \Omegac candidates and  $-0.15$ to $0.90\ps$ for the \Xicz candidates. Some candidates are reconstructed with negative decay times due to the finite decay time resolution; hence, the measurements include the negative decay time region. 
The normalisation mode candidates are analysed separately in two subsamples based on their lifetime, that match the ranges of the \Omegac and \Xicz lifetime measurements.


After the full selection sequence is applied, approximately one percent of the events contain more than one signal candidate. These multiple candidates are retained for the measurement~\cite{Koppenburg:2017zsh}, to avoid the risk of biasing the measurement in the choice of which candidates to keep.  


\section{Fits to the invariant-mass spectra}
\label{sec:massfit}

\begin{figure}[ht]
    \centering
    \includegraphics[width=0.49\textwidth]{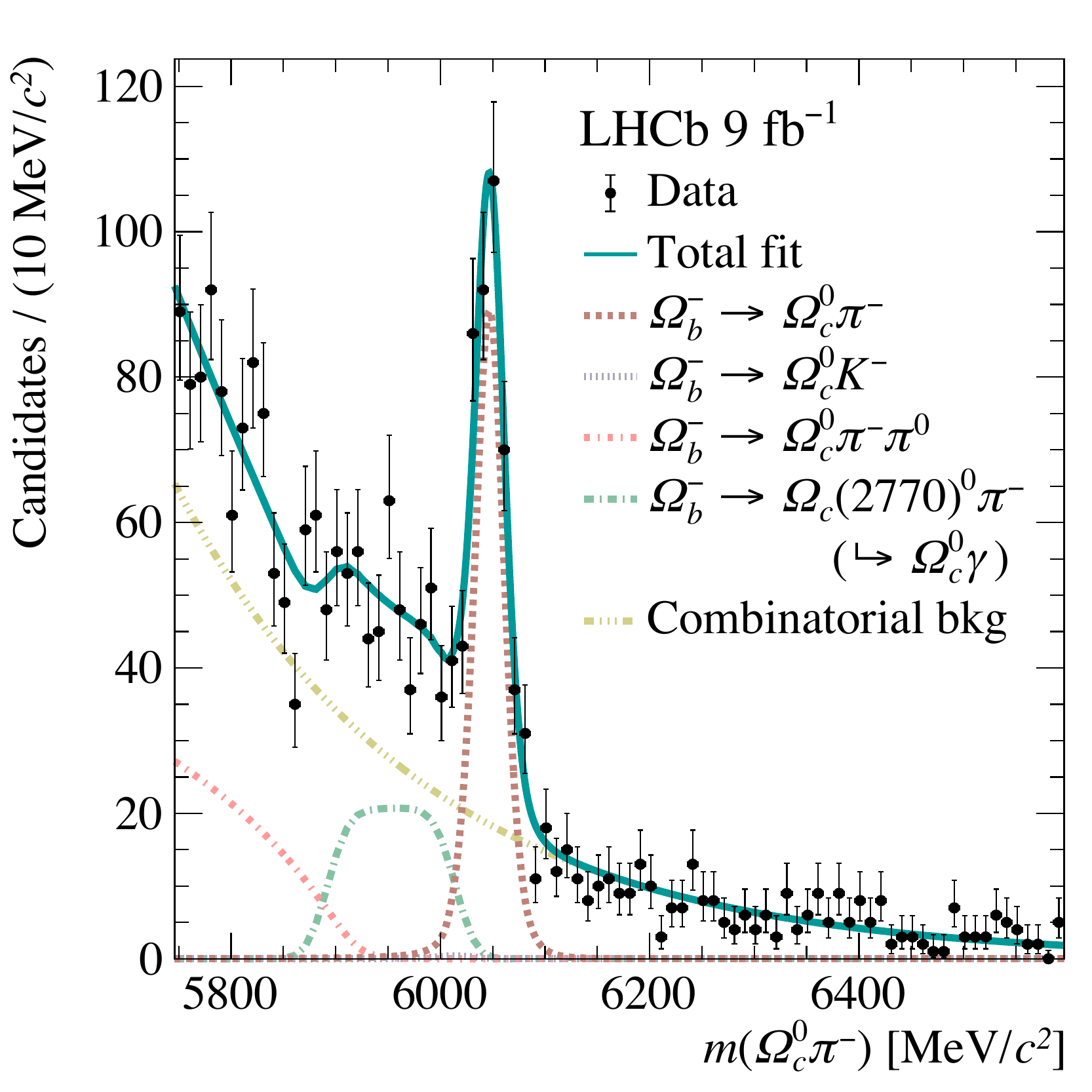}
    \includegraphics[width=0.49\textwidth]{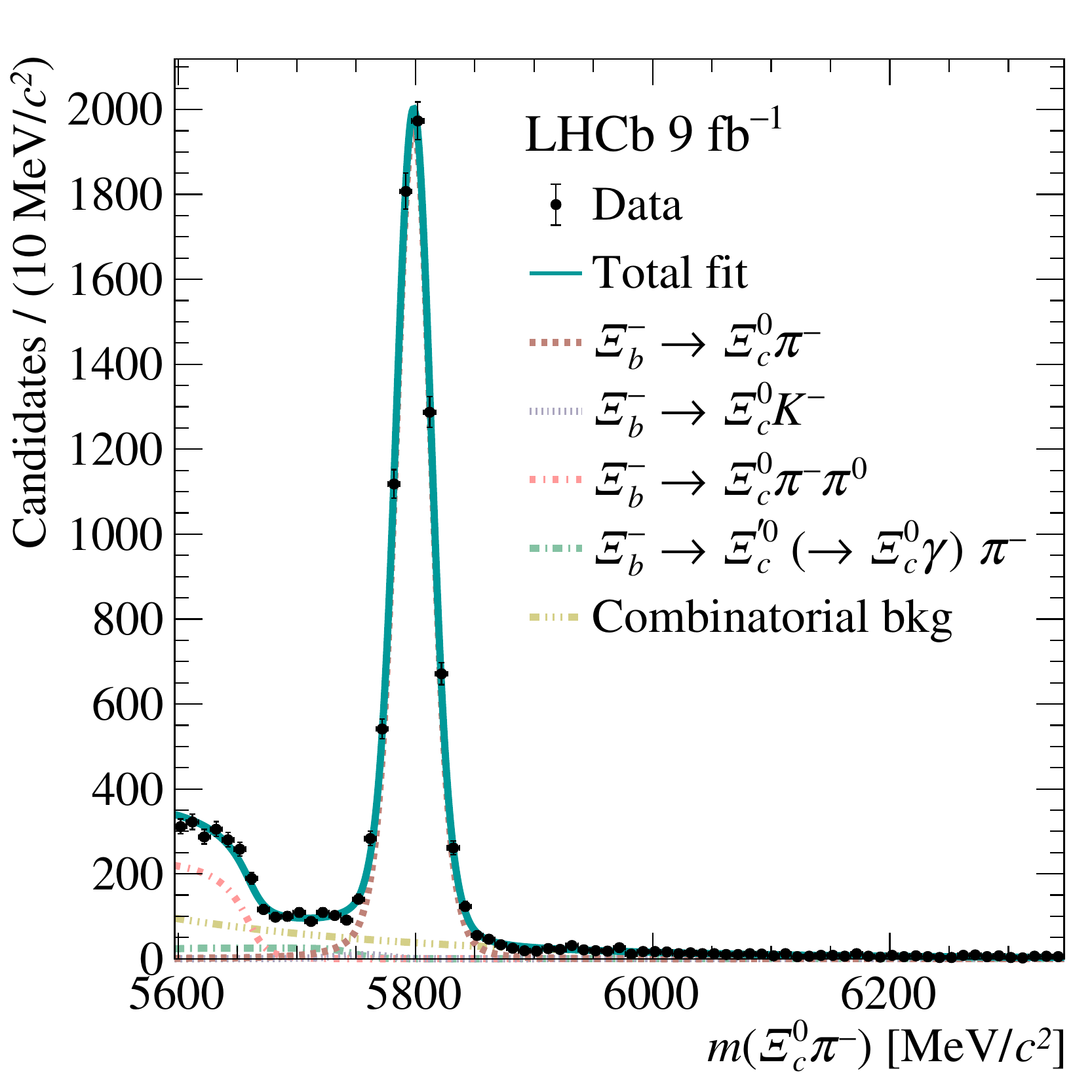}
    \includegraphics[width=0.49\textwidth]{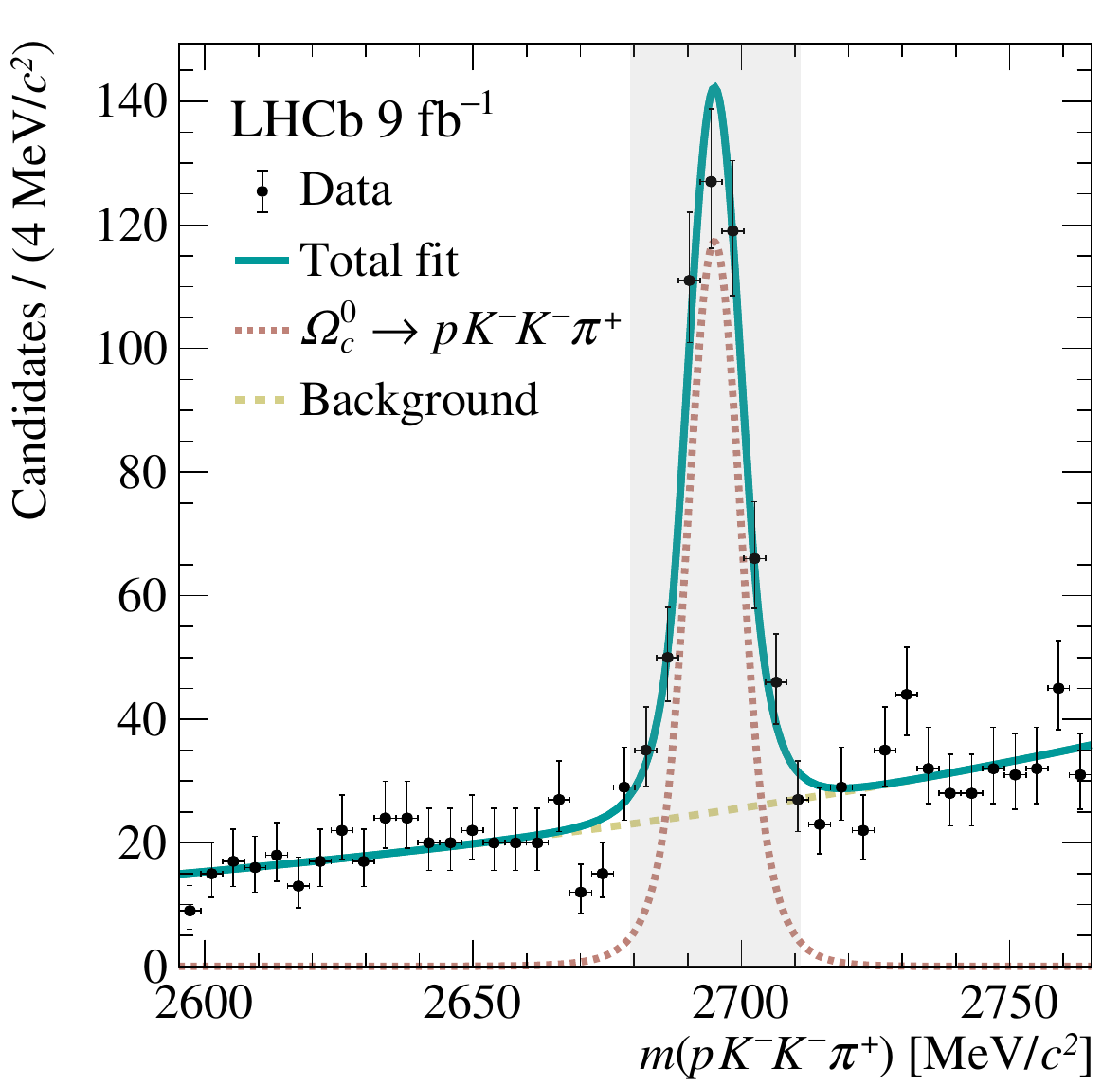}
    \includegraphics[width=0.49\textwidth]{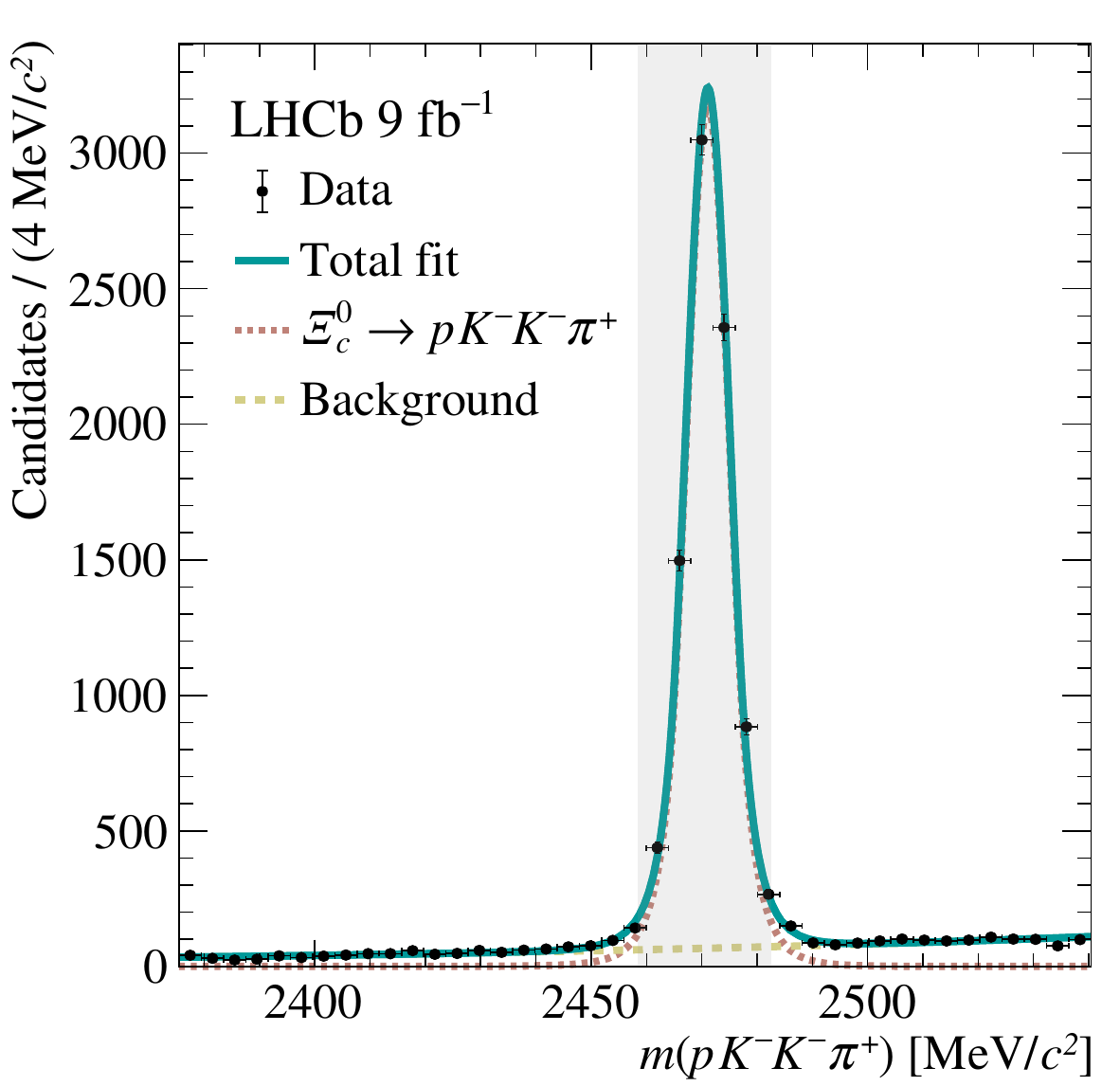}
    \caption{Top row: Invariant-mass distributions of reconstructed (left) $\Oms$ and (right) $\Xis$ candidates, with $\pKKpi$ invariant masses falling in the shaded regions of the bottom row plots. Bottom row: Distributions of the reconstructed $\pKKpi$ invariant mass, for (left)~$\Omegac$ and (right) $\Xicz$ candidates with invariant masses within three times the mass resolution of the $\Omegab$ and $\Xibm$ barons, respectively.}
    \label{fig:mass_fits}
\end{figure}

\begin{figure}[ht]
    \centering
    \includegraphics[width=0.49\textwidth]{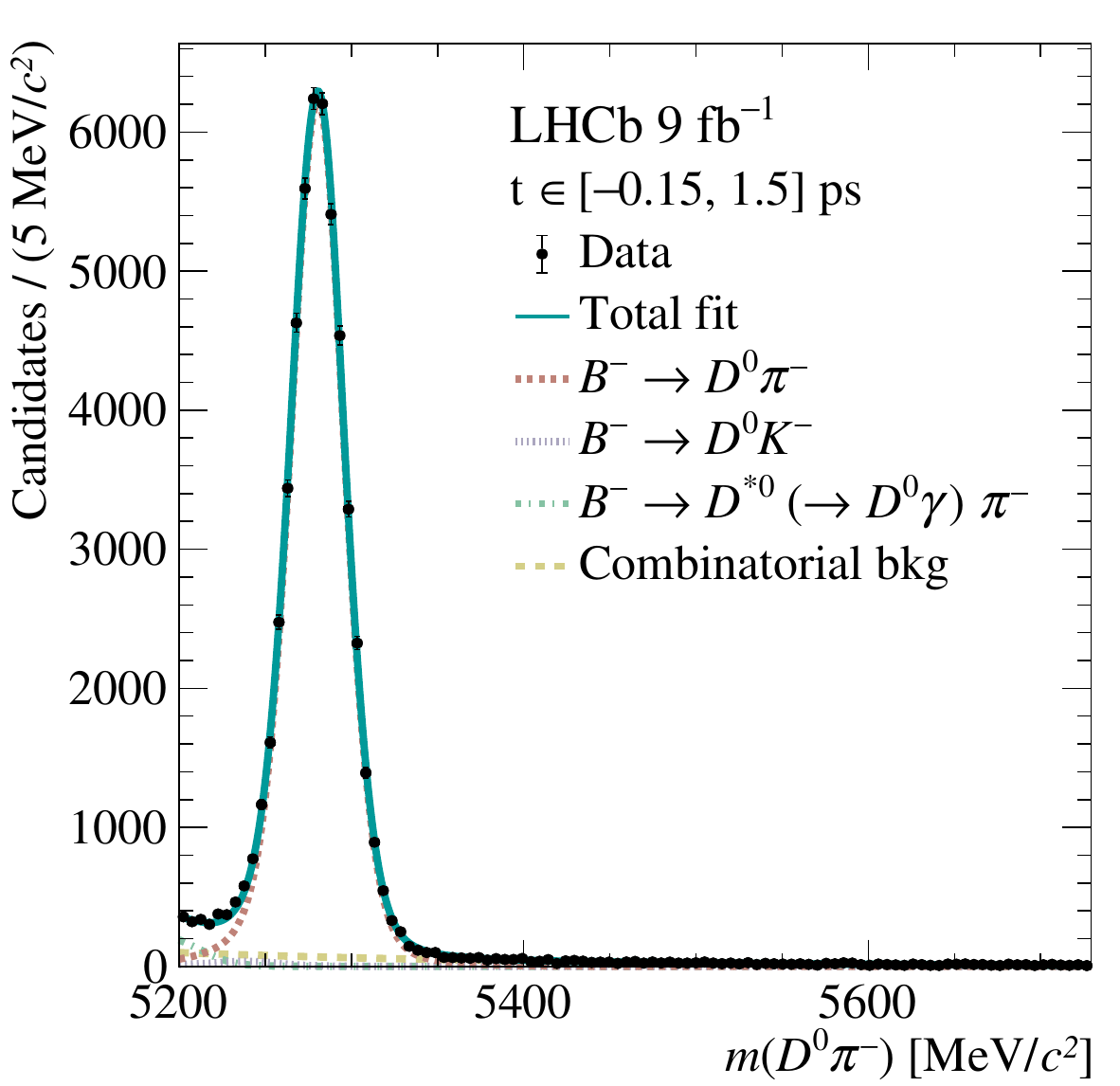}
    \includegraphics[width=0.49\textwidth]{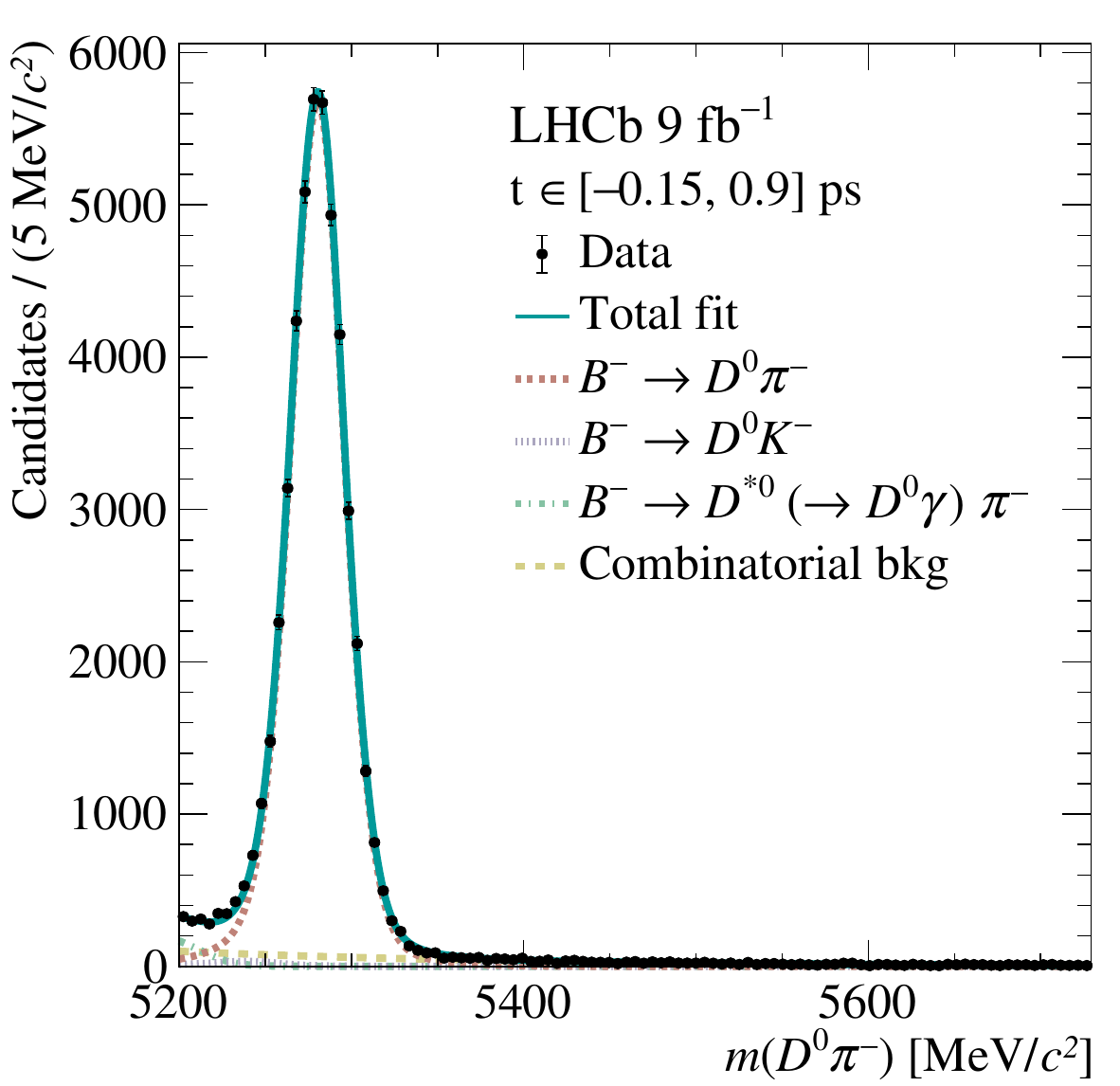}
    \includegraphics[width=0.49\textwidth]{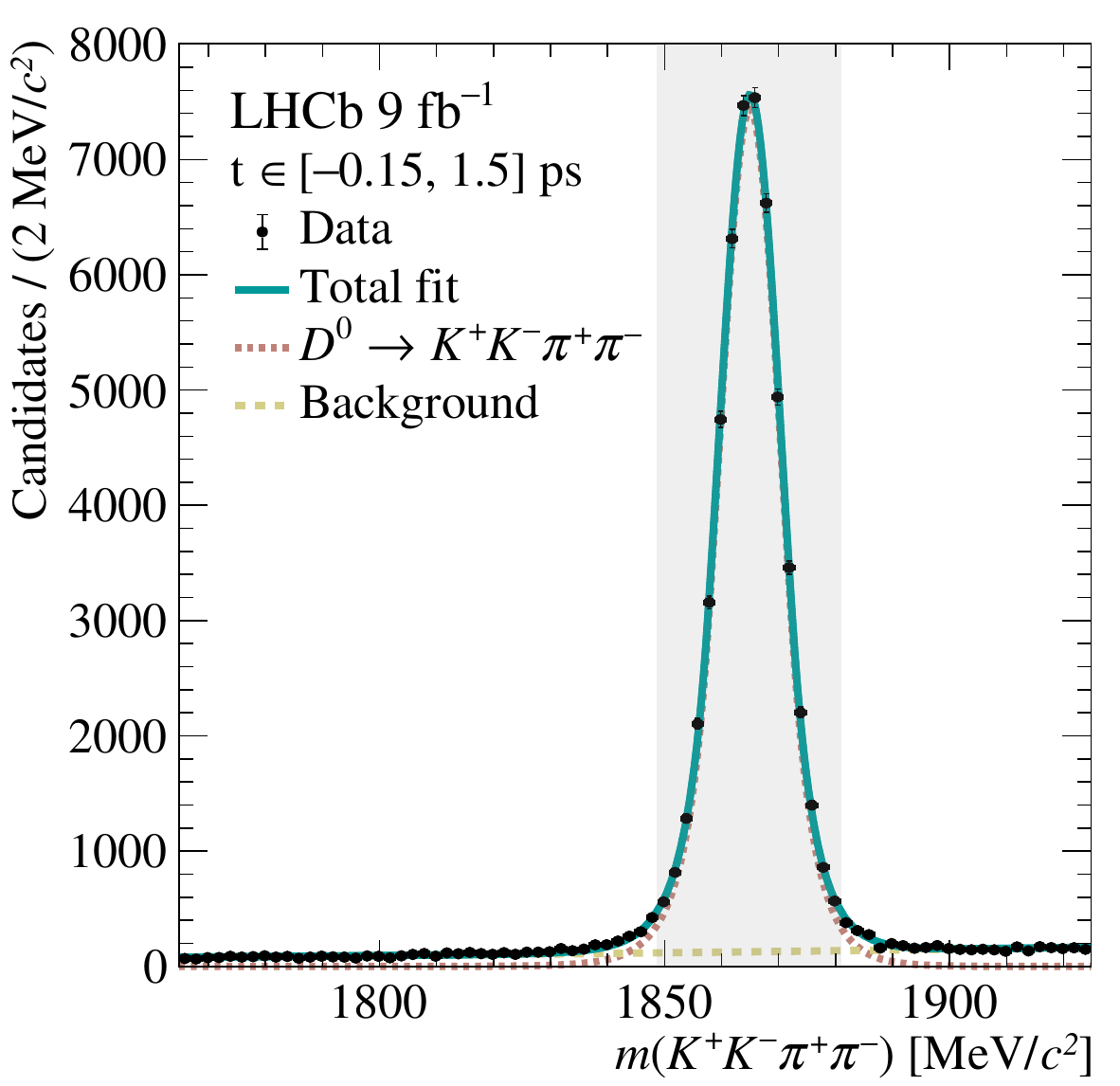}
    \includegraphics[width=0.49\textwidth]{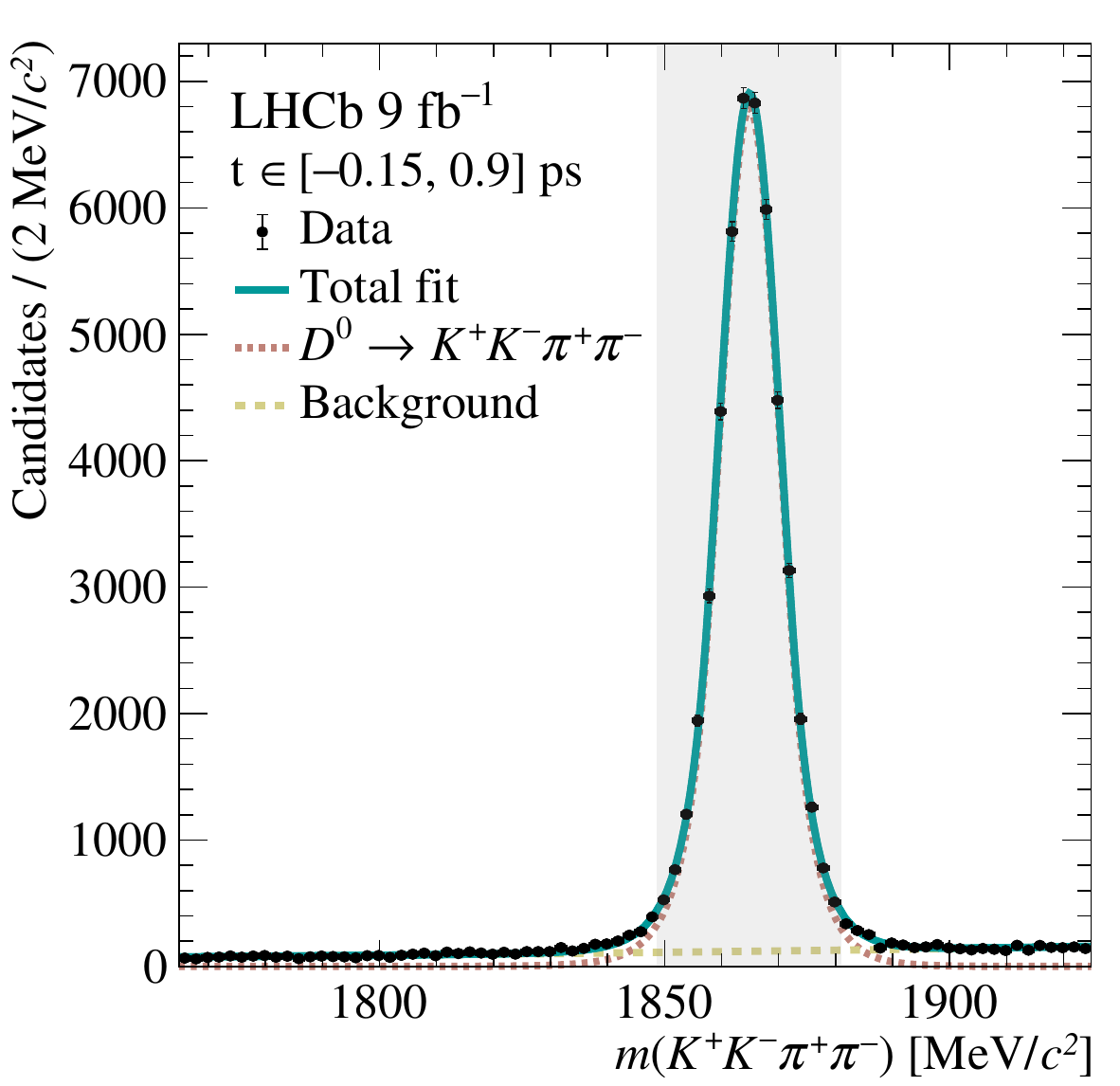}
    \caption{Top row: Invariant-mass distributions of reconstructed $\bees$ candidates, with $\KKpipi$ invariant masses falling in the shaded regions of the bottom-row plots. Bottom row: Distributions of the reconstructed $\KKpipi$ invariant mass for \Bub candidates with invariant masses within three times the mass resolution of the \Bub meson. Candidates with $\Dz$ decay times in the ranges used in the (left column) $\Omegac$ and (right column) $\Xicz$ lifetime measurements are included.}
    \label{fig:Bu_Dz_mass_fit}
\end{figure}

The signal and background components are separated through unbinned maximum-likelihood fits to the invariant-mass spectra of the selected candidates. The mass fits are performed to the \Oms, \Xis and \bees invariant-mass spectra, since this gives a better discrimination between the signal and backgrounds with charm hadrons in the decay chain than fits to the \pKKpi and \KKpipi spectra. The invariant-mass spectra and fits are shown in Fig.~\ref{fig:mass_fits} for the signal modes and in Fig.~\ref{fig:Bu_Dz_mass_fit} for the normalisation mode. 

The \bquark-hadron signals are modelled with a modified Crystal Ball function~\cite{Skwarnicki:1986xj} with power law tails on both sides of the peak (double-sided Crystal Ball, DCB), plus an additional Gaussian function. The tail parameters, as well as the relative widths and fractions of the two functions, are determined from simulation, whereas the peak position and overall width are determined from data.


The three spectra contain small fractions, less than 1\%, of the Cabibbo suppressed \OmsCS, \XisCS and \beesCS decays where the kaons have been misidentified as pions. These components are modelled with DCB functions where all parameters are determined from simulation and their yields are constrained relative to the yields of the signal decays. The branching fractions of the \bquark-baryon decays are not known; hence, the ratios of branching fractions between the Cabibbo favoured and suppressed decays are assumed to be the nominal Cabibbo suppression of approximately twenty, with an uncertainty of 30\% to account for \grpsuthree breaking effects. The branching fractions of the \bees and \beesCS decays are known with good accuracy and their ratio is calculated to be $(7.9\pm0.4)\%$~\cite{PDG2024}.
The relative efficiencies of the Cabibbo favoured and suppressed decays are determined from simulation. These ratios, including uncertainties, are combined to form Gaussian constraints in the fit.


Partially reconstructed decays, where one of the decay products is not included in the decay hypothesis, may enter the invariant-mass spectra. The decays \OmTostarfull, where the photons are not included, are modelled with a template determined from an analytic calculation. The shape depends on the helicity of the excited charm baryon, which is either $\frac{1}{2}$ or $\frac{3}{2}$, whose relative couplings are not known. Hence, both states are included as independent components of the fit, where their yields are determined to be approximately equal.


The corresponding partially reconstructed \XiToprime decays are  modelled with a sum of a generalised Gaussian function and a Johnsons's $S_U$ distribution~\cite{r:JohnsonSU}, where the parameters are determined from fast simulation~\cite{Cowan:2016tnm}. The dependence on the helicity of these decays is neglected since the yield is modest.  


The \Omcpi and \Xicpi invariant-mass spectra contain partially reconstructed \OmbToOmcpipizShort and \XibToXicpipizShort decays, where the \pim\piz either originate from $\rho(770)^-$ decays or are nonresonant. These decays are modelled with an Johnsons's $S_U$ distribution, with parameters fixed from fast simulation. The shapes of the invariant-mass distributions are not significantly affected by the helicity of the \pim\piz system~\cite{LHCb-PAPER-2021-012}. It is not possible to determine the relative fractions of the resonant and nonresonant decays from data; hence, the model giving smallest fit \chisq is selected. The nonresonant model is used for the fit to the \Omcpi invariant-mass spectrum and the resonant model is used for the \Xicpi invariant-mass spectrum. The yields of these decays are determined from data.


Partially reconstructed \bTogamma decays enter the \bees invariant-mass spectrum and are modelled with a Johnsons's $S_U$ distribution with parameters determined from fast simulation~\cite{Cowan:2016tnm}. The ratio of yields relative to the \bees decays, which is estimated from the relative branching fractions~\cite{PDG2024} and the relative reconstruction and selection efficiencies, is used as a Gaussian constraint in the fit.


Other peaking backgrounds are found to be negligible, for instance those from charmless \Xibm and \Omegab baryon decays.


The combinatorial background in these spectra originate from two sources: either from correctly reconstructed \Omegac, \Xicz or \Dz hadrons combined with an unassociated charged pion or from candidates built from five charged hadrons that originate from different processes in the event. The combinatorial background in the \Omcpi and \Xicpi invariant-mass spectra are dominated by the second category and the shapes are modelled with exponential functions, where the slope parameters are determined from a sample of candidates where the \pKKpi invariant mass is not consistent with originating from \Omegac or \Xicz decays. These parameters, including uncertainties, are used as Gaussian constraints in the fit to data.


The combinatorial background in the \bees spectrum is very small and modelled by an exponential function with an unconstrained slope.


The fits give signal yields of \mbox{$\yieldOmc$} for the \Oms decays and \mbox{$\yieldXicRound$} for the \Xis decays. The fitted yields for the \bees decays are \mbox{$\yieldDzOmRound$} (\mbox{$\yieldDzXiRound$}) in the decay time range used for the \Omegac (\Xicz) lifetime measurement.


\section{Lifetime measurements}
\label{sec:lifetime}


The \Omegac and \Xicz lifetimes are measured with a least-squares fit to the ratio of yields between the signal and normalisation modes, in bins of decay time. The bin boundaries are chosen to give approximately the same number of signal candidates in each bin, where six bins are used for the \Omegac lifetime measurement and fifteen are used for the \Xicz lifetime measurement. The ratio of yields in decay time bin $i$ with boundaries $t_i$  and $t_{i+1}$ is modelled to be
\begin{equation} 
    r_{i} \equiv \frac{\displaystyle\int_{t_i}^{t_{i+1}} A_{\Omegac,\Xicz}(t) \cdot \left(
    e^{-t/\tau_{\Omegac,\Xicz}} \ast R_{\Omegac,\Xicz}(t) \right)\deriv t}{\displaystyle\int_{t_i}^{t_{i+1}} A_{\Dz}(t) \cdot \left( e^{-t/\tau_{\Dz_{}}} \ast R_{\Dz}(t)\right)\deriv t},
    \label{eqn:fit_pdf}
\end{equation}
where $\tau_{\Omegac,\Xicz,\Dz}$ are the hadron lifetimes, $A_{\Omegac,\Xicz,\Dz}$ the decay time acceptance functions, $R_{\Omegac,\Xicz,\Dz}$ the decay time resolution functions and the symbol $\ast$ represents a convolution. The \Dz lifetime is fixed to its known value of $\tau_{\Dz} = 410.3\pm1.0\fs$~\cite{PDG2024}, while the charm baryon lifetimes are determined from the fit.


The decay time resolution function is determined from simulation and modelled with a sum of three Gaussian functions. The effective resolutions, defined as the square root of the sum of the Gaussian variances weighted by their fractions, are $83\fs$ for the \Omegac, $101\fs$ for the \Xicz and $75\fs$ for the \Dz decay times. A multiplicative scale factor to account for differences in resolution between data and simulation is determined from a fit to the \Dz decay time distribution. The distribution is modelled with an exponential function, with the \Dz lifetime fixed to its known value, convolved with the resolution function including a scale factor for its width, and multiplied by the acceptance function. The scale factor is determined to be $1.08\pm0.02$, which is applied to all three decay modes.


The decay time acceptance functions are determined from simulation, from the ratio between the reconstructed decay time distributions and a model consisting of an exponential function convolved with the resolution function. The lifetimes in the exponential functions are set to the values used in the simulation. The acceptance is almost independent of the charm hadron decay time, which is expected since the selection is designed to exploit the long \bquark-hadron lifetimes. They are modelled with the sum of second-order polynomials and a Gaussian function with a mean of approximately $-0.15\ps$ to account for a small deviation from a polynomial shape in that region.


Two different methods are employed to determine the decay time distribution of the signal. In the first method, the data are divided into bins of decay time and a mass fit is performed on each subset. The same fit functions as those described in Sec.~\ref{sec:massfit} are used, with only the yields allowed to vary and all other parameters fixed to the values given by the time-integrated mass fits. The ratios of signal and normalisation mode yields in each bin are fitted with the function given in Eq.~\ref{eqn:fit_pdf}.

The second method uses the \sPlot technique~\cite{Pivk:2004ty}, a statistical tool that determines the decay time distributions of the components included in the mass fits, using the \Oms, \Xis and \bees invariant masses as the discriminating variables. This results in binned decay time distributions for the signal and normalisation modes. The ratios of the distributions for the signal and normalisation modes are fit with the function given in Eq.~\ref{eqn:fit_pdf}.


Both methods return the generated lifetimes when applied to signal simulation, and give
unbiased results when evaluated on a large number of pseudoexperiments that include backgrounds. The statistical correlations of the measurements with the two methods, determined from pseudoexperiments, are 92\% for the \Omegac lifetime and 99\% for the \Xicz lifetime. The statistical and systematic uncertainties of the two methods are nearly identical. The results of the two methods were compared before unblinding and were found to be consistent when accounting for their total uncertainties and correlations. 

The first method for background subtraction, fitting the signal yields in bins of decay time, is used for the nominal result of this measurement. This method was selected since a similar method was used in a previous \lhcb measurement~\cite{LHCB-PAPER-2021-021} and since pseudoexperiments indicate that it is slightly more precise on average. This choice was made before the results were unblinded.

\begin{figure}[!t]
    \centering
    \includegraphics[width=0.49\textwidth]{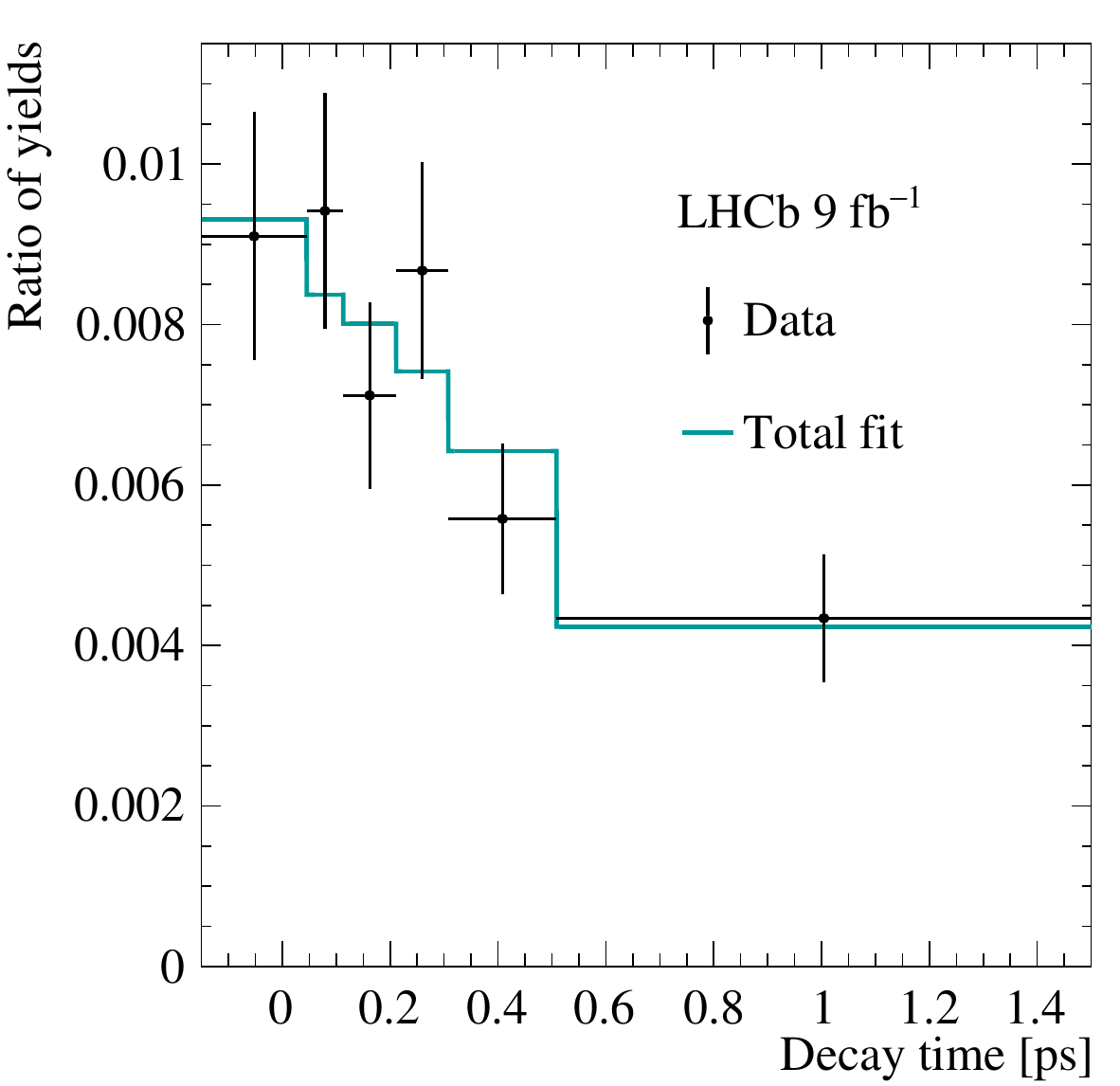}
    \includegraphics[width=0.49\textwidth]{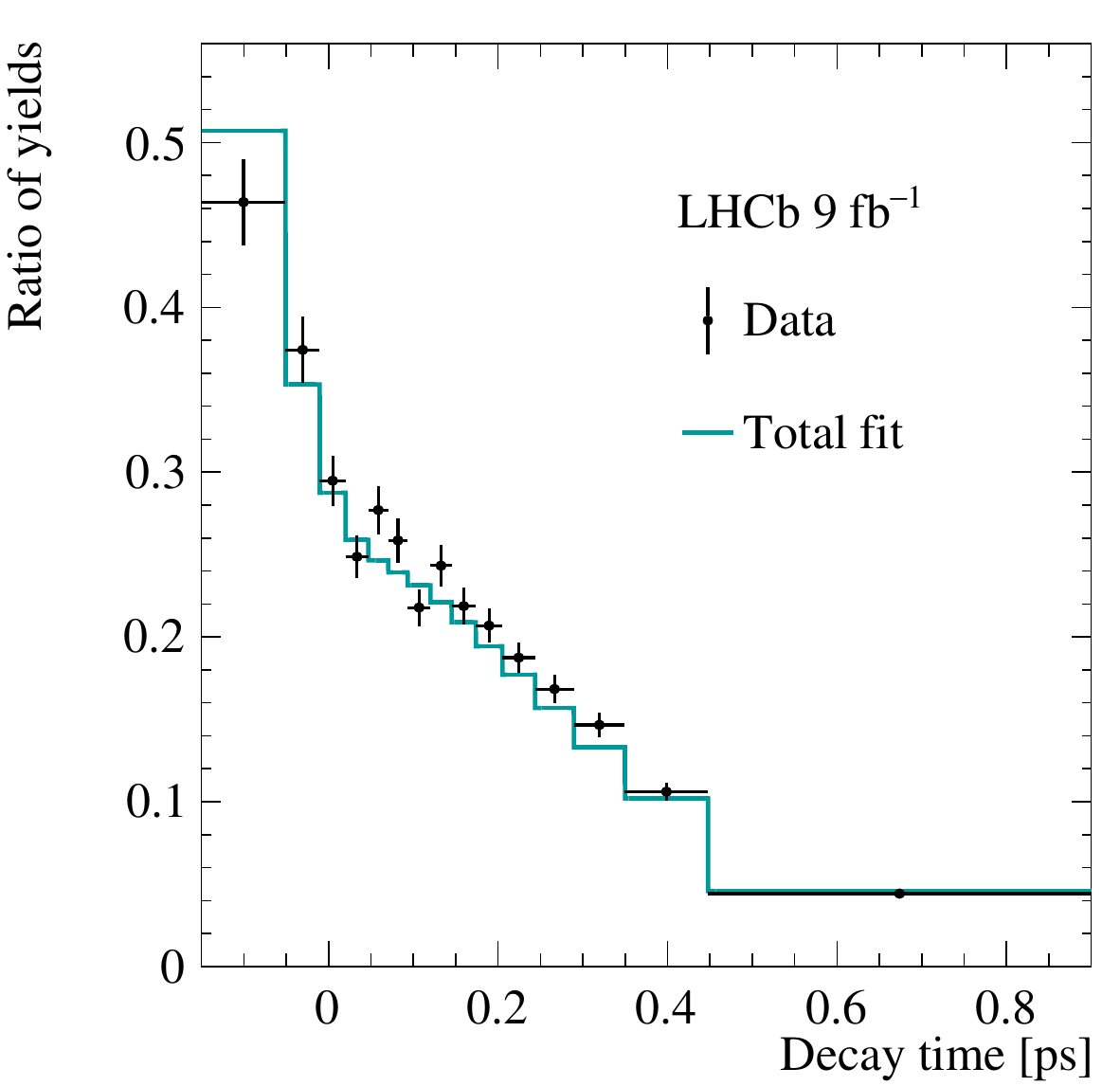}
    \caption{Ratio of yields ($r_{i}$) between the (left) $\Oms$ and (right) $\Xis$ decays, and the $\bees$ decays, in bins of decay time. The results from the fit with the model given in Eq.~\ref{eqn:fit_pdf} are shown.}
    \label{fig:lifetime_fit}
\end{figure}

The fits to the lifetime ratios are shown in Fig.~\ref{fig:lifetime_fit} and the measured lifetimes are 
\begin{eqnarray*}
    \tauOmc  &=& \systvalueOmStatRatioofYields\fs,\\
    \tauXicz &=& \systvalueXiStatRatioofYields\fs.  
\end{eqnarray*}

\noindent For comparison, the measured lifetimes for the second method are $\tauOmc = \systvalueOmStatSWeights$ and $\tauXicz = \systvalueXiStatSWeights$\fs.

\section{Systematic uncertainties}
\label{sec:systematics}


Several sources of systematic uncertainty are considered and they are summarised in Table~\ref{tab:syst}. In addition, the quoted statistical uncertainty includes a small component of systematic uncertainty due to the Gaussian constraints used in the mass fits.

\begin{table}[bt]
  \centering
  \caption{Systematic uncertainties for the \Omegac and \Xicz lifetime measurements.}
  \label{tab:syst}
  \begin{tabular}{l  cc}
    \toprule
    Sources    & $\sigma_{\tau_\Omegac}~\rm{[fs]}$ & $\sigma_{\tau_\Xicz}~\rm{[fs]}$ \\
    \midrule
    Signal mass modelling                          & \phantom{$<$} 0.1     & $<0.1$ \\
    Partially reconstructed background modelling   &             $<0.1$    & $<0.1$ \\
    Cabibbo suppressed \XiToOmczPim decay          &             $<0.1$    & \phantom{$<$} ---  \\
    Helicity of the \OmTostarfull decay            & \phantom{$<$} 0.3     & \phantom{$<$} ---  \\
    Combinatorial background modelling             & \phantom{$<$} 0.3     & $<0.1$ \\
    Decay time acceptance modelling                & \phantom{$<$} 1.7     & \phantom{$<$} 0.8 \\
    Decay time resolution modelling                & \phantom{$<$} 0.1     & \phantom{$<$} 0.5 \\
    Decay time resolution correction               &             $<0.1$    & $<0.1$ \\
    Uncertainty of the \Omegab and \Xibm lifetimes & \phantom{$<$} 0.5     & \phantom{$<$} 0.1  \\
    \midrule
    Total systematic uncertainty                   & \phantom{$<$} 1.8     & \phantom{$<$} 0.9 \\
    \midrule
    Uncertainty of the \Dz lifetime                & \phantom{$<$} 0.7     & \phantom{$<$} 0.4 \\
    \bottomrule
  \end{tabular}
\end{table}


The systematic uncertainties associated with the shape of the signal mass models are evaluated by generating pseudoexperiments with the nominal model described in Sec.~\ref{sec:massfit}, and fitting them with an alternative model consisting of a DCB function without the additional Gaussian function. The difference in average lifetime between the nominal and alternative fits, across all pseudoexperiments, is taken as the systematic uncertainty. 


The uncertainty related to the modelling of the partially reconstructed background is evaluated by generating pseudoexperiments using the nominal values for the parameters governing the shapes, and fitting them with parameter values that are varied within their uncertainties. The difference in fitted lifetime with the two sets of parameter values is determined for each pseudoexperiment, and the spread of this difference plus its average value is assigned as the systematic uncertainty. 
The decay \XiToXistar could contribute to the selected data, and an associated uncertainty is evaluated from pseudoexperiments that include this background, which are fit with the nominal model. This uncertainty is found to be negligible, since the background is absorbed in the modelling of the other partially reconstructed backgrounds.


Similarly, background from the Cabibbo suppressed decay \XiToOmczPim could be present in the \Oms mass spectrum, but its influence on the measurement is found to be negligible.


The uncertainty associated with the unknown fractions of the two helicity states for the \OmTostarfull decay is evaluated by generating pseudoexperiments with the helicity fractions taken from the nominal fit result. The pseudoexperiments are fit with two alternative models, either with helicity $\frac{1}{2}$ or $\frac{3}{2}$. The average of the difference between the nominal and alternative models is computed, and the largest of the two differences is taken as the systematic uncertainty.


The uncertainty related to the combinatorial background modelling is partially included in the statistical uncertainty, through the Gaussian constraint on the slope parameter. An additional uncertainty is assigned by fitting the pseudoexperiments with the slope parameter allowed to vary freely in the fit. The results from these fits are compared with those from the nominal fit, and the average difference is assigned as the systematic uncertainty. 


The systematic uncertainties associated with the decay time acceptance and decay time resolution modelling are evaluated by generating pseudoexperiments with the nominal models described in Sec.~\ref{sec:lifetime} and fitting them with alternative models. For the acceptance functions, a first order polynomial is used as the alternative model, without the additional Gaussian function. This is the dominant systematic uncertainty for both measurements.

The uncertainty related to the decay time resolution modelling is evaluated with an alternative model consisting of a sum of two Gaussian functions. The average differences between the nominal and alternative models are taken as the systematic uncertainties. 


The uncertainty on the decay time resolution scale factor extracted from data is propagated to the lifetime measurements. Pseudoexperiments are generated using the nominal value of the scale factor and fitted with the value varied within its uncertainty, independently for the signal and normalisation modes. The spread of this difference plus its average value is assigned as the systematic uncertainty.


The simulated samples used to determine the acceptance functions are generated with the known lifetimes for the \Omegab and \Xibm baryons~\cite{PDG2024}. Since these lifetimes are known with finite precision, the associated systematic uncertainty is evaluated by weighting the simulated samples to vary the \bquark-baryon lifetimes up and down by one unit of their uncertainty. The acceptance functions determined from these weighted samples are used as alternative models to generate pseudoexperiments, which are fit with the nominal model. The average shift in fitted lifetime is assigned as the systematic uncertainty.   


The charm baryon lifetimes are measured relative to the \Dz lifetime; hence, the fractional uncertainty on its lifetime is assigned as a separate systematic uncertainty.


The consistency of the results is validated by performing the measurement on subsets of the data, divided into years of data taking, polarity of the \lhcb magnet, centre-of-mass energy (7 and 8\tev versus 13\tev) and divided into two regions of $m(\Km\pip)$ invariant mass. The latter is to verify that the resonant structures present in the charm hadron decays do not affect the measured lifetimes. The data are also divided into two trigger categories, depending on if the hardware trigger decision was based on properties of the signal candidate or based on other particles produced in the $pp$ collisions. The results from all these subsets are consistent for both methods of background subtraction.


\section{Results and conclusion}
\label{sec:results}

The lifetimes of the \Omegac and \Xicz baryons are measured in fully reconstructed decay chains using data collected by the \lhcb experiment in the years 2011\,--\,2012 and 2015\,--\,2018 corresponding to an integrated luminosity of 9\invfb. The measured lifetimes are
\begin{eqnarray*}
\tauOmc \!\!&=&\!\! \systvalueOmroy\fs,\\
\tauXicz \!\!&=&\!\! \systvalueXiroy\fs,
\end{eqnarray*}
where the first uncertainty is statistical, the second is systematic and the third is due to the uncertainty of the \Dz lifetime. The statistical uncertainties are dominant, in particular for the \Omegac lifetime. These results are consistent with the previous measurements by the \lhcb experiment~\cite{LHCb-PAPER-2018-028,LHCb-PAPER-2019-008,LHCB-PAPER-2021-021}, confirming the revised hierarchy of charm baryon lifetimes, and provide guidance on the accuracy of the different techniques used to calculate charm baryon lifetimes. 


The measured lifetimes are combined with the previous \lhcb measurements~\cite{LHCb-PAPER-2018-028,LHCb-PAPER-2019-008,LHCB-PAPER-2021-021}, using a linear unbiased estimator~\cite{Nisius:2020jmf,Nisius:2014wua}, accounting for correlations. The measurements are statistically independent and all non-negligible systematic uncertainties are uncorrelated, apart from those related to the decay time resolution, which conservatively are considered to be fully correlated. 

The studies of semileptonic decays~\cite{LHCb-PAPER-2018-028,LHCb-PAPER-2019-008} measured the ratio between the charm baryons and the \Dp lifetimes, with the known value of the \Dp lifetime~\cite{PDG2018} as input. The knowledge of the \Dp lifetime has since improved~\cite{PhysRevLett.127.211801}, resulting in a new average~\cite{PDG2024} which is used to re-evaluate the  \Omegac and \Xicz lifetimes. The \lhcb combinations use these updated lifetimes.
The uncertainties associated with the \Dp and \Dz lifetime uncertainties are uncorrelated. 

The measurements that used promptly produced charm baryons~\cite{LHCB-PAPER-2021-021} were performed relative to a \Dz decay, using a previous average for the \Dz lifetime~\cite{PDG2020} as input. The measurements with hadronic decays use an updated average \Dz lifetime~\cite{PDG2024}, which includes a new measurement~\cite{PhysRevLett.127.211801}. The correlation between the systematic uncertainties associated with the \Dz lifetime between the two measurements is estimated to be 68\%.

The combinations give \chisqndf values of 0.13/2 for the \Omegac lifetime and 2.1/2 for the \Xicz lifetime and the resulting averages of the \lhcb measurements are
\begin{eqnarray*}
\tauOmc \!\!&=&\!\! \averageOmc\fs,\\
\tauXicz \!\!&=&\!\! \averageXic\fs,
\end{eqnarray*}
where the uncertainties include both statistical and systematic effects.

\section*{Acknowledgements}
%
%
\noindent We express our gratitude to our colleagues in the CERN
accelerator departments for the excellent performance of the LHC. We
thank the technical and administrative staff at the LHCb
institutes.
We acknowledge support from CERN and from the national agencies:
ARC (Australia);
CAPES, CNPq, FAPERJ and FINEP (Brazil); 
MOST and NSFC (China); 
CNRS/IN2P3 (France); 
BMBF, DFG and MPG (Germany); 
INFN (Italy); 
NWO (Netherlands); 
MNiSW and NCN (Poland); 
MCID/IFA (Romania); 
MICIU and AEI (Spain);
SNSF and SER (Switzerland); 
NASU (Ukraine); 
STFC (United Kingdom); 
DOE NP and NSF (USA).
We acknowledge the computing resources that are provided by ARDC (Australia), 
CBPF (Brazil),
CERN, 
IHEP and LZU (China),
IN2P3 (France), 
KIT and DESY (Germany), 
INFN (Italy), 
SURF (Netherlands),
Polish WLCG (Poland),
IFIN-HH (Romania), 
PIC (Spain), CSCS (Switzerland), 
and GridPP (United Kingdom).
We are indebted to the communities behind the multiple open-source
software packages on which we depend.
Individual groups or members have received support from
Key Research Program of Frontier Sciences of CAS, CAS PIFI, CAS CCEPP, 
Fundamental Research Funds for the Central Universities,  and Sci.\ \& Tech.\ Program of Guangzhou (China);
Minciencias (Colombia);
EPLANET, Marie Sk\l{}odowska-Curie Actions, ERC and NextGenerationEU (European Union);
A*MIDEX, ANR, IPhU and Labex P2IO, and R\'{e}gion Auvergne-Rh\^{o}ne-Alpes (France);
Alexander-von-Humboldt Foundation (Germany);
ICSC (Italy); 
Severo Ochoa and Mar\'ia de Maeztu Units of Excellence, GVA, XuntaGal, GENCAT, InTalent-Inditex and Prog.~Atracci\'on Talento CM (Spain);
SRC (Sweden);
the Leverhulme Trust, the Royal Society and UKRI (United Kingdom).




\addcontentsline{toc}{section}{References}
\bibliographystyle{LHCb}
\bibliography{main,standard,LHCb-PAPER,LHCb-CONF,LHCb-DP,LHCb-TDR}

\newpage
\centerline
{\large\bf LHCb collaboration}
\begin
{flushleft}
\small
R.~Aaij$^{38}$\lhcborcid{0000-0003-0533-1952},
A.S.W.~Abdelmotteleb$^{57}$\lhcborcid{0000-0001-7905-0542},
C.~Abellan~Beteta$^{51}$\lhcborcid{0009-0009-0869-6798},
F.~Abudin{\'e}n$^{57}$\lhcborcid{0000-0002-6737-3528},
T.~Ackernley$^{61}$\lhcborcid{0000-0002-5951-3498},
A. A. ~Adefisoye$^{69}$\lhcborcid{0000-0003-2448-1550},
B.~Adeva$^{47}$\lhcborcid{0000-0001-9756-3712},
M.~Adinolfi$^{55}$\lhcborcid{0000-0002-1326-1264},
P.~Adlarson$^{84}$\lhcborcid{0000-0001-6280-3851},
C.~Agapopoulou$^{14}$\lhcborcid{0000-0002-2368-0147}
% ...

{\footnotesize \it

% ...
$^{14}$Universit{\'e} Paris-Saclay, CNRS/IN2P3, IJCLab, Orsay, France\\
% ...
$^{38}$Nikhef National Institute for Subatomic Physics, Amsterdam, Netherlands\\
% ...
$^{47}$Instituto Galego de F{\'\i}sica de Altas Enerx{\'\i}as (IGFAE), Universidade de Santiago de Compostela, Santiago de Compostela, Spain\\
% ...
$^{51}$Physik-Institut, Universit{\"a}t Z{\"u}rich, Z{\"u}rich, Switzerland\\
% ...
$^{55}$H.H. Wills Physics Laboratory, University of Bristol, Bristol, United Kingdom\\
% ...
$^{57}$Department of Physics, University of Warwick, Coventry, United Kingdom\\
% ...
$^{60}$School of Physics and Astronomy, University of Glasgow, Glasgow, United Kingdom\\
$^{61}$Oliver Lodge Laboratory, University of Liverpool, Liverpool, United Kingdom\\
% ...
$^{69}$Syracuse University, Syracuse, NY, United States\\
% ...
$^{84}$Department of Physics and Astronomy, Uppsala University, Uppsala, Sweden, associated to $^{60}$\\
% ...
}
\end{flushleft}

\end{document}